\documentclass[]{new_aa2}  %
\usepackage{graphicx}

\begin{document}

\title{ The field population towards the globular cluster NGC 6553}

\author{Antonella Vallenari\inst{1} \and Sergio Ortolani\inst{2}  }

\offprints{A. Vallenari,
\email{vallenari@pd.astro.it}}

\institute{
Astronomical Observatory, Vicolo Osservatorio 5,
I-35122 Padova, Italy 
\and Dept. of Astronomy, Vicolo Osservatorio 5, I-35122 Padova, Italy\\
}

\date{Received September 2001 ; Accepted October 2001 }

\titlerunning{The Field population towards NGC 6553}

\authorrunning {A. Vallenari \& S.Ortolani}

\abstract {We analyze the colour magnitude diagram of the field population in the
direction of the bulge cluster NGC 6553 to discuss the stellar field
component projected towards the cluster. In particular
we investigate the nature of a peculiar feature, called
``red flare'' which appears as a faint turnoff, located  2 magnitudes about
below the turnoff of the cluster, with a sparse trace of a sub-giant branch
just above. A detailed modeling shows that the red flare is 
consistent with the expected location of 
the  turnoff and sub-giant branch of the bulge population, once that
the appropriate metallicity range and spatial distribution of the bulge
stars are adopted.  From the data, hints can be derived about the 
spatial shape and the basic stellar parameters of the bulge.
}

\maketitle

\section{ Introduction}

The structure of the Galactic Bulge and the nature of its stellar population 
have been  controversial 
problems in the last decade. Most of the current knowledge is based
on models derived from low 
resolution, photometric infrared observations.
The data interpretation requires some hypotheses on the distribution of the dust and some
assumption on the disk/bulge separation.
Efforts have been as well devoted to the interpretation of field 
color-magnitude diagrams (CMDs) projected in the direction of the inner bulge
along the minor and the major bulge axis. 
These diagrams appear quite complex, with a number of unexpected 
features. 
Recently Feltzing \& Gilmore  (\cite{felt}) and Beaulieu et al. (\cite{beau})
published HST CMDs of the globular cluster NGC 6553, projected at galactic
coordinates $l=5$ and $b=-3$.
In their diagrams they detect a peculiar feature called  ``red flare''.
This feature appears as a faint turnoff, located  2 magnitudes about
below the turnoff of the cluster, with a sparse trace of a sub-giant branch
just above.
The two papers leave  the problem open, suggesting different interpretations.
Feltzing \& Gilmore 
(\cite{felt}) suggest a possible contamination by the 
background  Sagittarius galaxy. Beaulieu et al. (\cite{beau})
seem more inclined to a bulge background turnoff. They fail to reproduce 
that turnoff with an isochrone of the same  [Fe/H] content than the cluster,
namely [Fe/H]$=-0.4$ concluding than the 
 mean bulge population must be more metal poor.
Both the interpretations, however, have
to face some difficulties: the Sagittarius galaxy hypothesis implies
a wide contamination, not present in the other fields,
while the ``more metal poor'' bulge field one 
is based on data where cluster and field stars are mixed together, preventing any firm 
conclusion. 
In addition Beaulieu et al. (\cite{beau})
reproduce the bulge  with a single metallicity stellar
population  located at 8 Kpc distance from the Sun,
while  recent papers based on high resolution spectroscopy
suggest a  wide range of [Fe/H],  from $-1.6$ to 0.55 with a mean
value of $-0.14$
(Rich \& McWilliam \cite{rich}). 
As far as the cluster is concerned, the  most recent literature
determinations, based on high resolution spectroscopy, give [Fe/H]$=-0.55\pm0.2$ or $Z\sim0.01$ (Barbuy \cite{barbu}),
 and [Fe/H]$=-0.16\pm0.1$ or $Z\sim0.02$ (Cohen et al. \cite{cohe}) 
revised  upward to [Fe/H]$\sim-0.06\pm0.15$ or $Z\sim0.027$
 by Carretta et al. (\cite{carr})
 with evidence of an excess of $\alpha-$process elements. 
From these results Cohen et al. (\cite{cohe}) and Carretta et al. (\cite{carr})
 conclude that the value of [Fe/H] for NGC 6553 is
comparable to the mean of the  Galactic Bulge. 
This implies that there is no strong evidence
in the recent literature supporting an higher metallicity
 of the cluster in comparison to the field.
For this reason we decide to proceed to a deeper investigation aimed
to detect the true nature of this feature using the galactic models
developed in Padova (Vallenari et al. \cite{valle2}) and the most recent CMDs of NGC 6553 published
by Zoccali et al. (\cite{zocc}) , where relative proper motions are provided. This allow us
to discard cluster objects from  field stars.
The goal is to discuss whether an appropriate admixture
of bulge and disk stars can be responsible
of this red flare.

In Section 2  the Galaxy model is presented, in Section 3 the observed and
simulated CMDs are shown and the resulting shapes of the bulge are
discussed. Finally the conclusions are drawn in Section 4.    

\section {Modeling the Galaxy}
\label{sec_model}
The simulation of the Galaxy is done with the code already described 
by Bertelli et al. (\cite{bert}) and revised by Vallenari et al. (\cite{valle2}).
The generation of the synthetic population makes use of the
set of stellar tracks by 
Salasnich et al. (\cite{sala})  for 
isochrones with $\alpha$ enhanced abundances.
We adopt for the old disk population an age of 10 Gyr
 while the youngest population is reaching an age  of 2 Gyr.

Evidence have been advanced in literature that the bulge might
not have spherical symmetry. 
 Several models have been proposed, going from spherical
to triaxial symmetry. In our galaxy model the following choices
are included:

1) COBE-DIRBE oblate spheroidal model (Dwek et al. \cite{dwek}, their model G0). It is
a Gaussian type function, with axial ratios 1:1:0.56, similar to the values
derived  on the basis of the near infrared maps 
 by Kent et al. (\cite{kent}).
 
2) COBE-DIRBE constrained triaxial model by Kent et al. (\cite{kent}) 
as modified by
Dwek et al. (\cite{dwek}) (their model G2). It is a ``boxy'' Gaussian models,
with axial ratios 1:0.22:0.16.

3) the triaxial
model introduced by Binney et al. (\cite{binn}) to fit COBE-DIRBE
data with the best fitting
parameters by  Bissantz et al. (\cite{biss}) (thereafter B0).
 The axial ratios are 1:0.6:0.4.

4) an exponential triaxial function (Dwek et al. \cite{dwek}  model E2)
 based on the
distribution of the IRAS Mira variable.
The axial ratios are 1:0.18:0.39. This model is suggested by Dwek et al. (\cite{dwek})
as giving one of the most reliable  fit to the COBE-DIRBE observations.

The angle between the sun-centre line and the major axis of the bar
is ranging from $25^\circ$ (Bissantz et al. \cite{biss}) to $20^\circ$ (Dwek et al. \cite{dwek}
 G2 model) while it becomes  $36^\circ$ in the case of the E2 model. 

A stellar population having ages from 12 to 10 Gyr and metal content
 ranging from $Z=0.008$ to 0.05 is adopted as suggested by
the most recent papers based on high resolution spectroscopy (Rich \& McWilliam \cite{rich},
 Barbuy et al. \cite{barb}).

\section{The red flare population in the field of NGC 6553}

The first step towards the interpretation of the data is the separation of 
the field from  the cluster stars.
Zoccali et al. (\cite{zocc}) compare  images of NGC~6553 taken at different epochs
in order to derive star proper motions. As expected, these authors 
find that the cluster and the field  populations follow two different distributions
 of relative 
proper motions  and can  consequently be disentangled. 
Cluster objects are distributed by construction in a Gaussian centered on relative proper motion 
in $l$ and $b$
 $(dl,db)=(0,0)$ while field population is clustered around 
$(dl,db)=(-0.245,-0.017)$ pixels.
Following the analysis by Zoccali et al. (\cite{zocc}), 
in their CMD of the region around NGC~6553 (see Fig.\ref{redd_clu})
 we select  stars having $dl<-0.2$ (corresponding to a proper
motion of $\mu_l<5.89$ mas yr$^{-1}$ of the cluster with respect
to the field objects).
Those stars are believed to sample the foreground and background population.
Consequently,  the field  population is disentangled from the cluster objects.
The corresponding CMD is presented in Fig.\ref{buliso}. About 1200 stars
are selected in the field.
In particular the red flare is brought into evidence. 
Since NGC~6553 is located in the direction of the Galactic bulge, 
the most obvious hypothesis is that this population is simply due to
a superposition along the line of sight of mainly bulge stars with
a residual disk contamination. The stars redder than $(V-I)\sim2.15$
would then be 
sub-giant and clump objects of disk and bulge.
In order to reproduce the red flare under this hypothesis, it is necessary to remember that a 
single isochrone
at a fixed distance of 8 Kpc is
not an appropriate representation of the bulge population. In fact the bulge
is  an admixture of objects of various metallicity, and  
has a spatial distribution as discussed in previous Sections. 
Fig. \ref{buldist} presents the predicted distributions of the bulge stars
with the distance from the Sun in the
line of sight of NGC~6553 at varying bulge models.
The vast majority of the  models presented in literature 
expect bulge stars to be located at a distance from 5 to 11 Kpc from the Sun.
Fig. \ref{buliso} where isochrones are superposed on red flare data
shows how populations of various metallicity placed 
at different distance from the Sun can reproduce the location and the colour
of the sub-giant branch and clump.

A  Galactic disk component is needed
to reproduce the main sequence region  bluer than $(V-I)\sim 1.8$
and roughly brighter than $V=20.5$ mag, as far
as the appropriate reddening is adopted (see the following Section).
The evolved stars of this disk population are located in the 
same CMD  region of the sub-giant branch and clump stars of the bulge.

\subsection {Extinction determination}
\begin{figure}
\resizebox{\hsize}{!}{\includegraphics{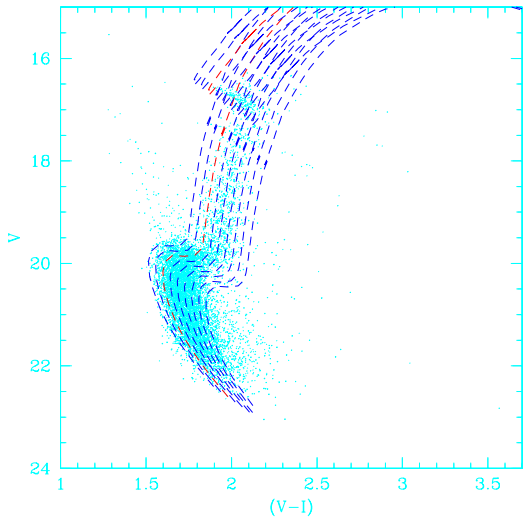}}
\caption{The observational CM diagram of the cluster NGC 6553 from
Zoccali et al.  (\cite{zocc}). Superposed on it a grid of isochrones at varying reddening  from
  $A_V=2.0$ to 2.8 in steps of 0.1.
  The age is  12 Gyr, the metal content  $Z=0.019$, with  enhanced abundances of $\alpha$ elements   
in respect to Solar (Salasnich et al. \cite{sala}). 
The assumed distance modulus is $(m-M)_0=13.6$ as derived
by Zoccali et al. \cite{zocc}.  }
\label{redd_clu}
\end{figure}

It has been proved (Vallenari et al. \cite{valle2}) that the blue edge  of the main
 sequence in the CMD of the  disk and bulge population can be interpreted in terms
of extinction along the line of sight. Comparing  the observed main sequence
edge location with theoretical simulations
 at varying extinction,  
 we derive the values of  $A_V$ along the 
line of sight as a function of the heliocentric distance
(see Fig.\ref{buliso}).
 Using this method,
we find that
at the distance of 5 Kpc from the Sun  the absorption is 
 $A_V=2.7$ mag. This value is  larger than  $A_V=2.08$
 obtained by Zoccali et al. (\cite{zocc}) 
from the fit of the cluster NGC 6553, located at 5.3 Kpc from us.
 However, these authors claim they analyze a low reddening region inside the cluster. 
In fact, differential reddening is present in the field. This is 
 evident from Fig.\ref{redd_clu} where cluster data of the whole field obtained by 
Zoccali et al. (\cite{zocc}) are plotted
together with isochrones at varying  extinction.
If the width of the cluster sub-giant branch is mainly due to differential reddening,
the data are consistent with $A_V$ ranging
 from
2.1 to 2.7. This result is in substantial agreement with  
the analysis made by 
 Sagar et al. (\cite{saga}), Guarnieri et
al. (\cite{guar}), Beaulieu et al.  (\cite{beau}) who find a differential reddening of about 0.5 mag inside the 
field.
As far as the foreground and background population of Fig.\ref{buliso} is
concerned,
 the main contribution to the absorption is due to the disk
interstellar gas and dust at distances closer than 5-6 Kpc. In fact,
 to reproduce the colour of the main sequence 
field population at higher distances, 
an additional extinction of 0.1 is required leading to a final value of
$A_V=2.8$.

\subsection{The results}
\begin{figure}
\resizebox{\hsize}{!}{\includegraphics{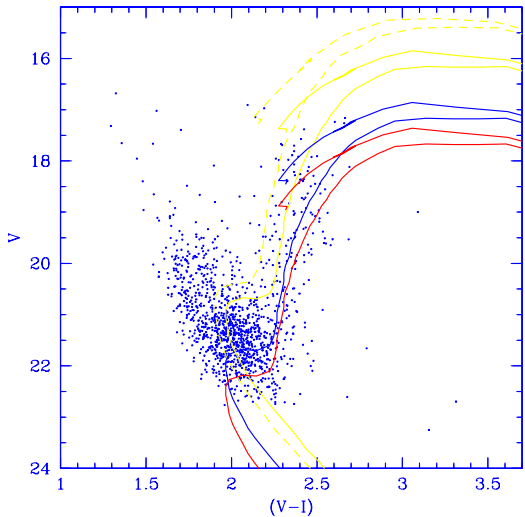}}
\caption{The observational CM diagram of the field population
towards NGC 6553 together with three isochrones of $Z=0.019$, age  12 Gyr,
 and
distance of 5, 8,10 Kpc (solid lines), and an isochrone of $Z=0.008$, age 12 Gyr
and distance of 5 Kpc (dotted line)}
\label{buliso}

\end{figure}

\begin{figure}
\resizebox{\hsize}{!}{\includegraphics{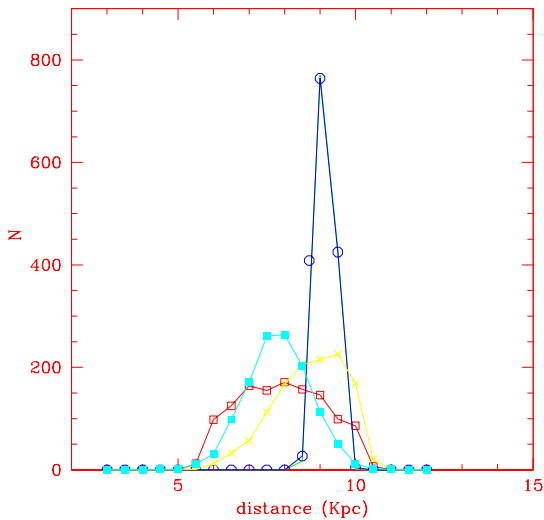}}
\caption{The expected distribution of bulge stars with the distance from 
the Sun at varying spatial distributions in the line of sight of NGC~6553.
 Open 
squares indicate the model by Bissantz et al.  (\cite{biss}), circles show the model E2, crosses refer to the 
model G2 and finally filled squares represent the model G0 
(see text for detail).}
\label{buldist}

\end{figure}

\begin{figure}
\resizebox{\hsize}{!}{\includegraphics{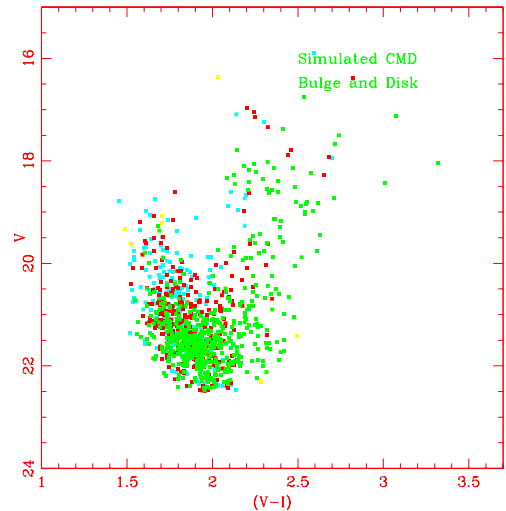}}
\caption{The simulated CM diagram of the background and foreground population
towards NGC 6553. This simulation includes bulge and disk stars.}
\label{bulsim1}

\end{figure}

Using the model described in  Section~\ref{sec_model} including
thin disk and bulge we reproduce the CMD of Fig.\ref{buliso}.
The result of the simulation is shown in Fig.\ref{bulsim1}.
It is quite obvious looking at Fig.\ref{buldist} that different spatial
distributions of the stars in the bulge can predict  different
turnoff magnitudes.  
The termination point magnitude  changes  at maximum of 0.3-0.4 mag because
of this effect. A detailed discussion can be found
in  Vallenari et al.  (\cite{valle1}). Here we recall that
``rounder'' distributions (model G0 or B0) result in brighter magnitudes
($\sim 20.4$),
while flat distributions having large position angle (model E2) produce
fainter turnoff ($\sim 20.7$).  
The luminosity function of the disk population reaches a 
maximum at $V \sim 20.5$ mag corresponding to  a distance from us of about 
7~kpc. 
 Because of the combined effect of disk and bulge population,
to compare 
 observational and simulated CMDs is then not trivial. We make use of  a $\chi^2$ test 
(see Vallenari et al. \cite{valle2} for more detail
on the adopted procedure).
$\chi^2$ is defined as
$\chi^2 = 1/\nu \sum _i (a_i-b_i)^2/a_i$ where $a_i$ and $b_i$
are the observed and the expected number of stars per magnitude bin,
and $\nu$ is the number of degrees of freedom (which is equal to the number
of bins minus 1, since the in the simulations we impose that the
total number of model stars is equal to the observed total number of stars).
In consequence of this definition, high probability models have $\chi^2 \sim 1$.
Due to the poor statistics, all the models,
but the model E2
are in good agreement with the data, resulting in a $\chi^2 \sim 1.5-2$.
Although it is difficult to reach a firm conclusion, models B0 and G0 seem
to result in a more convincing agreement with the data, model G2 is
not in disagreement, while
the model E2 is definitely less probable ($\chi^2 \sim 3$).
As a final comment on the adopted bulge population,
we point out  that while the metallicity
and distance spread 
are fundamental to reproduce the observational width of the sub-giant
branch, the age range is poorly constrained by the data, due to
the poor statistics.
This analysis strongly points in favor of the fact that
 the red flare is simply due 
to the  bulge and disk populations.
From our simulations, we estimate that
the disk contributes at maximum for the 30\% to the red sub-giant branch
and clump population.

\section {Conclusions}

Using simulations derived from  models of the Galaxy and the recent 
HST CMDs of NGC 6553 and its projected field, we are able to reproduce the
bulge and disk field contributions and the so-called ``red flare''
discussed in Feltzing \& Gilmore  (\cite{felt}) and Beaulieu et al.  (\cite{beau}).
This feature turns out to be a natural
consequence of  bulge and disk 
field contamination.
The spread in distance and metallicity is fundamental to reproduce the
observational CMD of the field.
 The bulge is modeled using  an age from 10 to 12 Gyr and a
metallicity ranging between $Z=0.008$ and $Z=0.05$. The cluster is fitted with
an isochrone of $Z=0.019$ which is inside the metallicity range of the bulge field.  
We recall however that while constraints set on metallicity
are stringent, the age range 
is poorly determined, due to the limited statistics.

\end{document}